\documentclass[aps,prb,twocolumn]{revtex4}
\usepackage{graphicx}

\begin{document}

\title{Transition of stoichiometric Sr$_2$VO$_{3}$FeAs to a superconducting state at 37.2 K}

\author{Xiyu Zhu, Fei Han, Gang Mu, Peng Cheng, Bing Shen, Bin Zeng, and Hai-Hu Wen}\email{hhwen@aphy.iphy.ac.cn }

\affiliation{National Laboratory for Superconductivity, Institute of
Physics and Beijing National Laboratory for Condensed Matter
Physics, Chinese Academy of Sciences, P. O. Box 603, Beijing 100190,
China}

\begin{abstract}
The superconductor Sr$_4$V$_2$O$_6$Fe$_2$As$_2$ with transition
temperature at 37.2 K has been fabricated. It has a layered
structure with the space group of \emph{p4/nmm}, and with the
lattice constants a = 3.9296$\AA$ and c = 15.6732$\AA$. The observed
large diamagnetization signal and zero-resistance demonstrated the
bulk superconductivity. The broadening of resistive transition was
measured under different magnetic fields leading to the discovery of
a rather high upper critical field. The results also suggest a large
vortex liquid region which reflects high anisotropy of the system.
The Hall effect measurements revealed dominantly electron-like
charge carriers in this material. The superconductivity in the
present system may be induced by oxygen deficiency or the multiple
valence states of vanadium.

\end{abstract} \maketitle

Since the discovery of superconductivity\cite{1} at 26 K in
oxy-arsenide $LaFeAsO_{1-x}F_x$, tremendous attention has been paid
to searching new superconductors in this family. Among the
superconductors with several different structures,\cite{2,3,4,5,6}
the highest T$_c$ has been raised to 55-56 K\cite{7,8,9,10,11} in
doped oxy-iron-arsenides (F-doped LnFeAsO, the so-called 1111 phase,
Ln=rare earth elements) or the fluoride derivative iron-arsenides
(Ln-doped AEFeAsF, AE=alkaline earth elements).\cite{ChengPEPL} The
superconductivity can also be induced by applying a high pressure to
the undoped parent samples.\cite{13,14} Although it remains unclear
what governs the mechanism of superconductivity in the FeAs-based
system, it turns out to be clear that the parent phase is
accompanied by an antiferromagnetic (AF) order and the
superconductivity can be induced by suppressing this magnetic order.
A typical example was illustrated in the $(Ba,Sr)Fe_2As_2$
(so-called 122) system, the AF order is suppressed and
superconductivity was induced by either doping K to the Ba or Sr
sites,\cite{2,Canfield,ChuCW2} or doping Co to the Fe
sites.\cite{Mandrus,Fisher} On the other hand, superconductivity was
also found in the parent phase of FeP-based system, such as LaFePO
(T$_c$ = 2.75K)\cite{FeP}, or in LiFeAs.\cite{3,4} Very recently
superconductivity at about 17 K was found in another FeP based
parent compound Sr$_4$Sc$_2$O$_6$Fe$_2$P$_2$ (so-called
42622).\cite{FeP42622} Due to the absence of the AF order in the
superconductors mentioned above, one naturally questions whether the
AF order is a prerequisite for the superconductivity in the
iron-pnictide system. As far as we know, no superconductivity was
detected in the parent phase of some FeAs-based compounds, including
the 1111, 122 and the recently discovered 42622 and 32522
phases.\cite{FeAs32522,GFChen,XYZhu2,Ogino,Johrendt} Although some
trace of superconductivity was reported in the doped FeAs-based
42622 or 32522 compounds, the high-$T_c$ superconductivity was not
supported by a clear large diamagnetization
signal.\cite{GFChen,XYZhu2} In this Letter, we report the discovery
of superconductivity at about 37.2 K in the new compound
Sr$_4$V$_2$O$_6$Fe$_2$As$_2$. This work presents the unambiguous
evidence for high temperature superconductivity in the FeAs-based
42622 system.

The polycrystalline samples were synthesized by using a two-step
solid state reaction method.\cite{ZhuXY} Firstly, SrAs powders were
obtained by the chemical reaction method with Sr pieces and As
grains. Then they were mixed with V$_2$O$_5$ (purity 99.9\%), SrO
(purity 99\%), Fe and Sr powders (purity 99.9\%), in the formula
Sr$_4$V$_2$O$_6$Fe$_2$As$_2$, ground and pressed into a pellet
shape. The weighing, mixing and pressing processes were performed in
a glove box with a protective argon atmosphere (the H$_2$O and O$_2$
contents were both below 0.1 PPM). The pellets were sealed in a
silica tube with 0.2 bar of Ar gas and followed by a heat treatment
at 1150 $^o$C for 40 hours. Then it was cooled down slowly to room
temperature. The X-ray diffraction (XRD) patterns of our samples
were carried out by a $Mac$-$Science$ MXP18A-HF equipment with
$\theta - 2\theta$ scan.  The XRD data taken using powder sample was
analyzed by the Rietveld fitting method using the GSAS
suite\cite{GSAS}. The dc susceptibility of the samples were measured
on a superconducting quantum interference device (Quantum Design,
SQUID, MPMS-7T). The resistivity and Hall effect measurements were
done using a six-probe technique on the Quantum Design instrument
physical property measurement system (PPMS) with magnetic fields up
to 9 T. The temperature stabilization was better than 0.1\% and the
resolution of the voltmeter was better than 10 nV.

The X-ray diffraction (XRD) pattern for the sample
Sr$_4$V$_2$O$_6$Fe$_2$As$_2$ is shown in Fig. 1. This compound
consists of a stacking of anti-fluorite Fe$_2$As$_2$ layers and
perovskite-type Sr$_4$V$_2$O$_6$ layers. Rietveld refinement shown
by the solid line in the figure gives good agreement between the
data and the calculated profiles. The impurity phase was detected
and found to come mainly from Sr$_2$VO$_4$. The calculated XRD
pattern (solid line) of the mixture was obtained by adopting a ratio
Sr$_4$V$_2$O$_6$Fe$_2$As$_2$ : Sr$_2$VO$_4$ = 13 : 1. Lattice
parameters for the tetragonal unit cell was determined to be a =
3.9296$\AA$ and c = 15.6732$\AA$. In Table I, the structure
parameters were listed with agreement factors: wR$_{p}$ = 13.23$\%$,
R$_{p}$= 9.35$\%$.

\begin{table} \caption{Fitting parameters for
Sr$_4$V$_2$O$_6$Fe$_2$As$_2$. wR$_{p}$ = 13.23$\%$, R$_{p}$=
9.35$\%$.}
\begin{tabular}{cccccccccc}
\hline \hline
Atom & site & x & y & z  & \\
\hline
 V   &  2 c &0.2500        &  0.2500   &  0.3081 &\\
Fe  &  2 a & 0.2500      &      0.7500   & 0.0000 &\\
 Sr  &  2 c &   0.7500    &   0.7500    &   0.1903  &\\
 Sr  &  2 c &     0.7500  &    0.7500   & 0.4145&\\
As  &  2 c &     0.2500   &   0.2500      &  0.0909  &\\
 O   &  4 f &   0.2500    &     0.7500    &  0.2922 & \\
 O   &  2 c &     0.2500    &    0.2500    & 0.4318   &\\

 \hline \hline

\end{tabular}
\label{tab:table1}
\end{table}

To confirm the presence of bulk superconductivity in our sample, we
measured the magnetization of our sample using the dc susceptibility
method. In Fig. 2(a) we present the temperature dependent dc
susceptibility data measured with a dc field of 10 Oe. The data were
obtained using the zero-field-cooling and field-cooling modes. The
onset superconducting transition temperature as determined from the
dc magnetization is about 31.5 K. In Fig. 2(b), we show the
temperature dependence of resistivity under zero field in the
temperature region up to 300 K. A clear superconducting transition
can be observed in the low temperature region. The onset critical
transition temperature was determined to be about 37.2 K from this
curve and the resistivity drops to zero at the temperature of about
31 K, being rather consistent with the onset transition point in the
dc susceptibility curve and further confirming the bulk
superconductivity in this compound. A metallic behavior can be seen
above the transition temperature. Interestingly the resistivity in
the normal state exhibits a huge plateau-like shape in the high
temperature region. This could be attributed to the incomplete
suppression to the possible AF order (if it exists for this
compound), or it is similar to that in the hole doped
FeAs-superconductors $Ba_{1-x}K_xFe_2As_2$\cite{2} and
$(La,Pr)_{1-x}Sr_xFeAsO$\cite{WenEPL,MuGPRBPrSr} where a general
feature of bending down of resistivity was observed in the high
temperature region. This will be clarified by future experiment. If
the former case is true, the superconducting transition temperature
can be further increased by adding electrons or holes into the
sample.

\begin{figure}
\includegraphics[width=9cm]{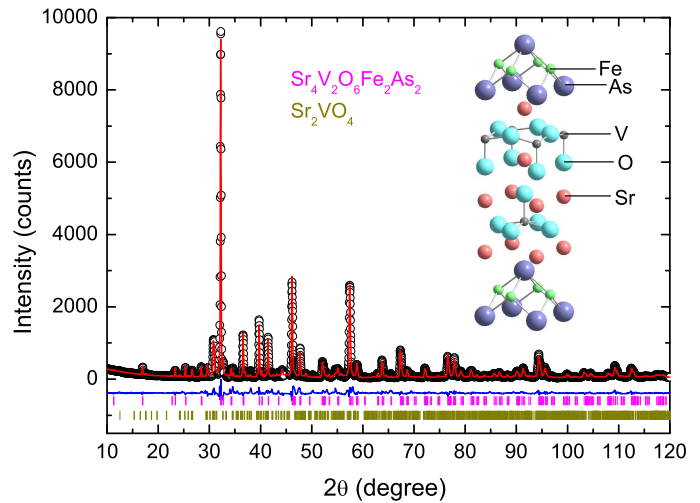}
\caption{(Color online) X-ray diffraction patterns for the sample
Sr$_4$V$_2$O$_6$Fe$_2$As$_2$. Bottom: Sr$_2$VO$_4$ was included in
the refinement as the minor impurity phase.} \label{fig1}
\end{figure}

\begin{figure}
\includegraphics[width=9cm]{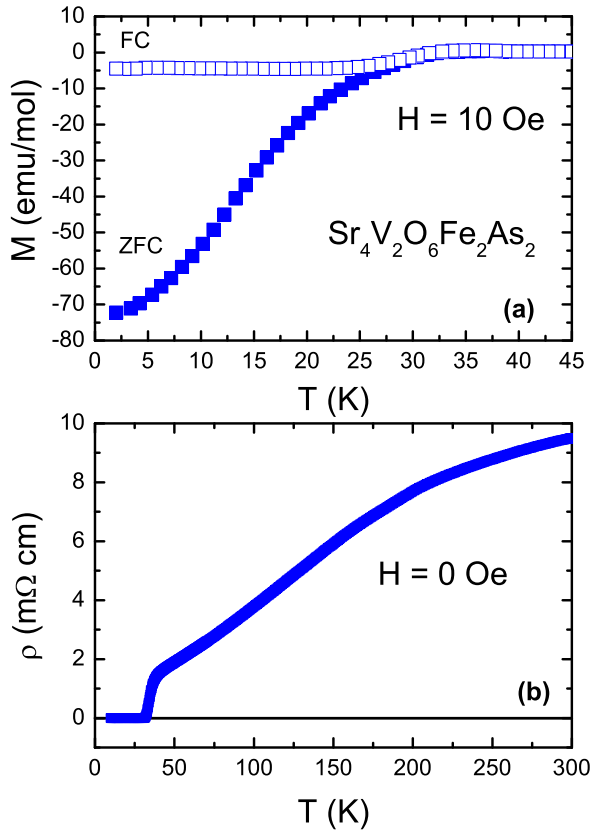}
\caption{(Color online) (a) Temperature dependence of the dc
susceptibility for the sample Sr$_4$V$_2$O$_6$Fe$_2$As$_2$. The dc
susceptibility data were obtained using the zero-field-cooling and
field-cooling modes with a dc mganetic field of 10 Oe. The onset
superconducting transition temperature was determined to be 31.5 K.
(b) Temperature dependence of the resistivity under zero field in
temperature region up to 300 K.} \label{fig2}
\end{figure}

We present the resistivity data in low temperature region under
different fields in Fig. 3(a). The transition curve is clearly
rounded near the onset transition temperature, showing the
possibility of the presence of superconducting fluctuation in this
system. This is actually understandable since the system now becomes
more 2D-like due to the very large spacing distance between the FeAs
planes. One can see that the onset transition temperature moves very
little at a field as high as 9 T. While the zero-resistance
temperature moves to low temperatures rapidly, showing a broadening
effect induced by the magnetic field which may imply the presence of
superconducting weak-link between the grains in the present sample.
We have pointed out that the evolution of onset transition
temperature with field mainly reflects the information of upper
critical field along the ab-plane for a polycrystalline
sample.\cite{ZhuXY} Therefore we took a criterion of 90\%$\rho_n$ to
determine the onset critical temperatures under differen fields.
Surprisingly, we got a rather large slope of $H_{c2}$(T) near $T_c$,
$(dH_{c2}/dT)_{T=T_c} \approx$ -11.3 T/K. This value is obviously
larger than that obtained in other FeAs-based
superconductors,\cite{MuGPRBPrSr,ZhuXY,ZSWang} and consequently
results in a rather high upper critical field of about 302 T using
the Werthamer-Helfand-Hohenberg (WHH) formula $H_{c2}(0)=-0.69T_c
dH_{c2}/(dT)_{T=T_c}$.\cite{WHH} In Fig.3 we present also the
irreversibility line $H_{irr}$ taking with the criterion of 0.1\%
$\rho_n$. It is found that a large region exists between the upper
critical field $H_{c2}(T)$ and the irreversibility field
$H_{irr}(T)$. As mentioned before, this large separation may be
induced by the weak link effect between the grains. In this sense
the superconducting coherence length (most probably along the
c-axis) is shorter in this material than in other families, like
1111 and 122. Furthermore, this large gap between $H_{c2}(T)$ and
$H_{irr}(T)$ can also be explained by the stronger thermal
fluctuation effect of vortices in the present system due to the
higher anisotropy.

\begin{figure}
\includegraphics[width=9cm]{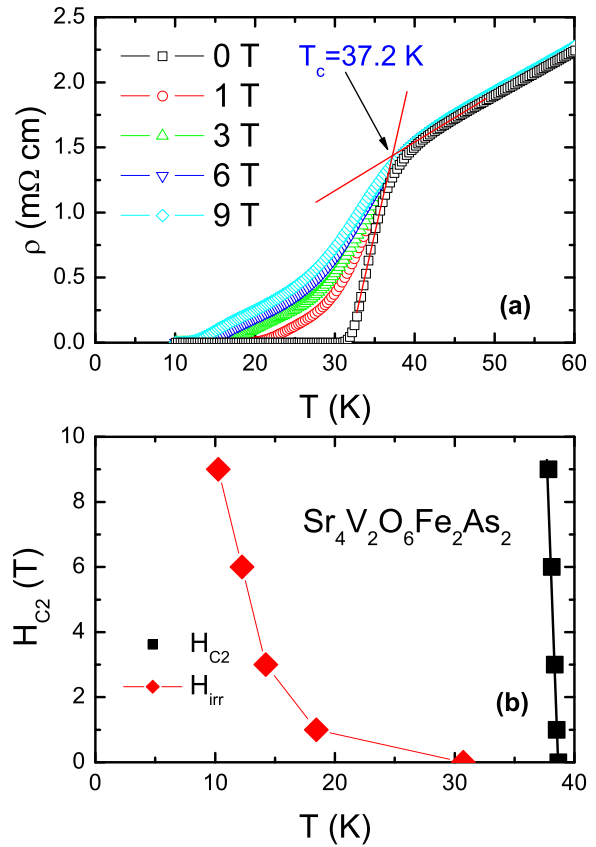}
\caption{(Color online) (a) Temperature dependence of resistivity in
the low temperature region under different fields. The onset
transition temperature was determined to be 37.2 K in the data under
zero field. (b) The phase diagram plotted as $H_{c2}$ versus T. A
criterion of 90\% $\rho_n$ was taken to determine the upper critical
fields. The irreversibility line $H_{irr}$ taking with the criterion
of 0.1\% $\rho_n$ is also presented in this figure.} \label{fig3}
\end{figure}
\begin{figure}
\includegraphics[width=9cm]{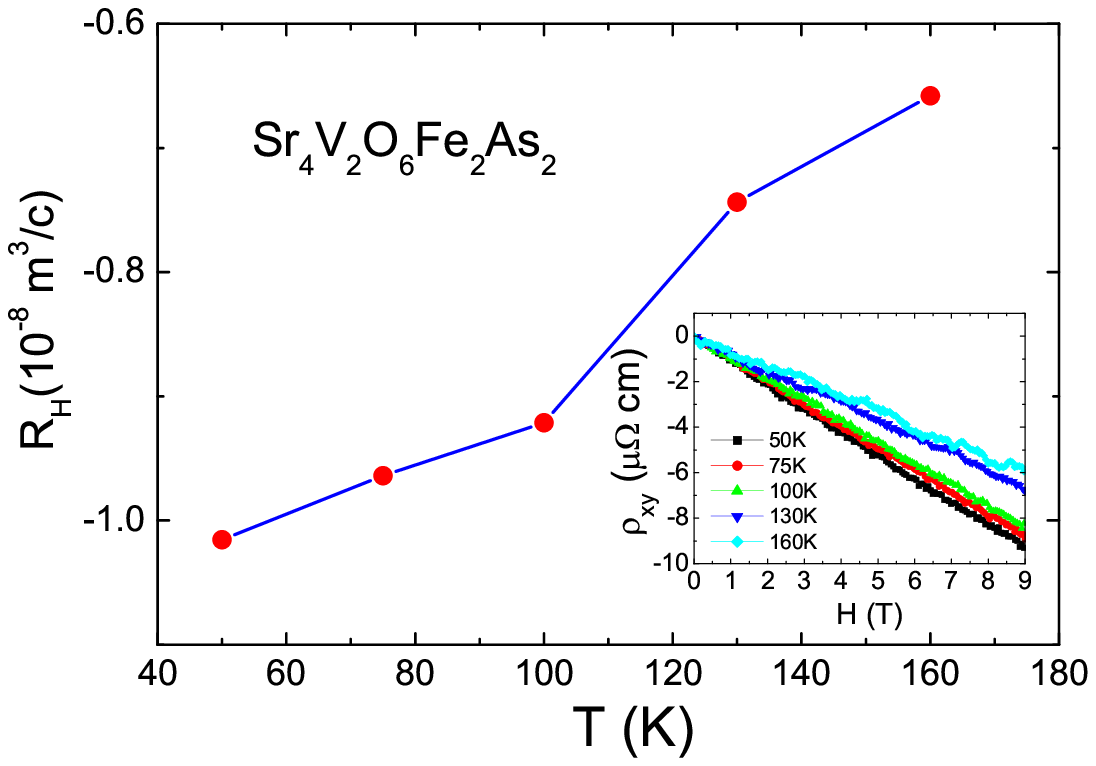}
\caption{(Color online) Temperature dependence of Hall coefficient
$R_H(T)$ determined at 9 T from the transverse resistivity
$\rho_{xy}$ shown in the inset. It is clear that the transverse
resistivity $\rho_{xy}$ is negative and linearly related to the
magnetic field, suggesting that the electron conduction is dominated
by the electron-like charge carriers in
Sr$_4$V$_2$O$_6$Fe$_2$As$_2$.} \label{fig4}
\end{figure}

In order to know the electronic properties of this parent phase, we
measured the Hall effect in the normal state. Fig.4 shows the
temperature dependence of the Hall coefficient R$_H(T)$. As shown in
the inset, the raw data of the transverse resistivity $\rho_{xy}$ is
negative and exhibits a linear relation with the magnetic field.
This is similar to that in other FeAs-based
superconductors.\cite{ZhuXY} The Hall coefficient R$_H$ is negative
in the measured temperature region indicating that the electron-like
charge carriers are dominating the conduction. However, as in all
other FeAs-based superconductors, R$_H$(T) shows a strong
temperature dependence, which is actually anticipated by the
multiband picture: the electron scattering rate $1/\tau_i$ of each
band will vary with temperature in a different way, therefore a
combined contribution of multiple bands will lead to strong
temperature dependence of Hall coefficient.\cite{YangHuanPRL2008}

As stated previously, without additional doping, thus called as
parent phase of FeP-based materials\cite{FeP,FeP42622} and
LiFeAs\cite{3,4} show superconductivity, but with relatively low
$T_c$. So far there has been no report about the existence of the
static long range AF order in the parent or doped phase of FeP-based
compound and in LiFeAs. In the FeAs-based parent phase, however, in
most cases (with an exception of LiFeAs\cite{3,4} and FeAs-32522
parent phase)\cite{FeAs32522} an AF order was observed in the low
temperature region and the superconductivity can only be achieved by
suppressing this unique AF order. Recently through careful Hall
effect measurements and analysis, it was concluded that the AF order
and superconductivity actually compete each other for the
quasiparticle density of states in the underdoped
$Ba(Fe_{1-x}Co_x)_2As_2$\cite{FangL}. In the present work the
superconductivity was observed in the undoped phase of
Sr$_4$V$_2$O$_6$Fe$_2$As$_2$, and this raises the question again
whether the AF order is a prerequisite for the superconductivity.
One possibility for explaining the superconductivity in this system
is that the oxygen content in the sample may be tunable since there
are many occupying sites for oxygen atoms in the structure. Oxygen
deficiency in the system implies a doping of electrons and thus
leads to the superconductivity. If this is true, doping to this
"parent" phase will lead to the suppression of superconductivity and
make the AF order emerge. Another possibility may be the multiple
valence states of vanadium. For example, in the compound V$_2$O$_3$,
the vanadium has a "3+" valence state, while that in Sr$_2$VO$_4$ is
"4+". Therefore our compound here provides a new platform for the
doping and tuning the superconductivity and AF magnetism. In
addition, there may be other possibilities to explain the
superconductivity in this system. For example, the present system
shares the similarity of the FeP-based and LiFeAs parent compounds
in which no evidence of AF order was found. Our results will call
for band structural calculations for the detailed structure of the
Fermi surface for this new compound. A naive understanding would
assume that the FeAs-planes are very similar to that in other
systems, thus it has no reason for the absence of the AF order in
the present system if it is induced by the nesting effect. Since
this is the first observation of superconductivity (with a
relatively high transition temperature) in 42622 phase of the
FeAs-based compound, our discovery will stimulate the in-depth
understanding to the mechanism of superconductivity in the iron
pnictide superconductors.

In summary, superconductivity with $T_c$ = 37.2 K was found in the
undoped phase of FeAs-based compound Sr$_4$V$_2$O$_6$Fe$_2$As$_2$.
The x-ray diffraction measurement showed that this compound has a
rather pure phase exhibiting a layered structure and the space group
of \emph{P4/nmm}. Both the large diamagnetization signal and
zero-resistance were detected, indicating an unambiguous evidence
for bulk superconductivity. The broadening of resistive transition
was measured under different magnetic fields and the upper critical
field (most possibly the H$_{c2}^{ab}$) determined by using the
Werthamer-Helfand-Hohenberg (WHH) formula is as high as 302 T. The
Hall effect measurements showed that the conduction in this material
was dominated by the electron-like charge carriers. This is the
first report of superconductivity with high-$T_c$ in the undoped
phase of the FeAs-based 42622 family. Based on this material
platform, more new superconductors, with probably higher T$_c$ are
expectable.

This work was supported by the Natural Science Foundation of China,
the Ministry of Science and Technology of China (973 Projects
No.2006CB601000, No. 2006CB921802), and Chinese Academy of Sciences
(Project ITSNEM).


\begin{thebibliography}{00}
\bibitem{1}Y. Kamihara, T. Watanabe, M. Hirano, and H. Hosono, J. Am. Chem. Soc. \textbf{130}, 3296 (2008).
\bibitem{2}M. Rotter, M. Tegel, and D. Johrendt, Phys. Rev. Lett. \textbf{101}, 107006 (2008).
\bibitem{3}X. C. Wang, Q. Q. Liu, Y. X. Lv, W. B. Gao, L. X. Yang, R. C. Yu, F. Y. Li, and C. Q. Jin, Solid State Commun. \textbf{148} 538 (2008).
\bibitem{4} J. H. Tapp, Z. Tang, B. Lv, K. Sasmal, B. Lorenz, C. W. Chu, and A. M. Guloy, Phys. Rev. B \textbf{78}, 060505(R) (2008).
\bibitem{5} F. C. Hsu, J. Y. Luo, K. W. Yeh, T. K. Chen, T. W. Huang, P. M. Wu, Y. C. Lee, Y. L. Huang, Y. Y. Chu, D. C. Yan, and M. K. Wu,
Proc. Natl. Acad. Sci. 105, 14262-4 (2008).
\bibitem{6}T. Klimczuk, T. M. McQueen, A. J. Williams, Q. Huang, F.Ronning, E. D. Bauer, J. D. Thompson, M. A. Green, and R. J. Cava, Phys.
Rev. B \textbf{79}, 012505 (2009).
\bibitem{7} X. H. Chen, T. Wu, G. Wu, R. H. Liu, H. Chen, and D. F. Fang, Nature \textbf{453},761 (2008).
\bibitem{8} Z. A. Ren, W. Lu, J. Yang, W. Yi, X. L. Shen, Z. C. Li, G. C. Che, X. L. Dong, L. L. Sun, F. Zhou, and Z. X. Zhao, Chin. Phys. Lett. \textbf{25}, 2215 (2008).
\bibitem{9} Y. Aiura, K. Sato, H. Iwasawa, Y. Nakashima, A. Ino, M. Arita, K. Shimada, H. Namatame, M. Taniguchi, I. Hase, K. Miyazawa, P. M. Shirage, H. Eisaki, H. Kito, A. Iyo, J. Phys. Soc. Jpn. \textbf{77}, 103712 (2008).

\bibitem{10}P. Cheng, L. Fang, H. Yang, X. Zhu,
G. Mu, H. Q. Luo, Z. S. Wang, and H. H. Wen, Science in China G.\textbf{51}, 719-722 (2008).

\bibitem{11} C. Wang, L. Li, S. Chi, Z. Zhu, Z. Ren, Y. Li, Y. Wang, X. Lin, Y. Luo, S. Jiang, X. Xu, G. Cao, and Z. Xu, Europhys. Lett. \textbf{83}, 67006 (2008).
\bibitem{ChengPEPL}P. Cheng, B. Shen, G. Mu, X. Zhu, F. Han, B. Zeng, and H. H. Wen , Europhys. Lett. \textbf{85}, 67003 (2009).
\bibitem{13} H. Okada, K. Igawa, H. Takahashi, Y. Kamihara, M. Hirano, H. Hosono, K. Matsubayashi, Y.
Uwatoko, J. Phys. Soc. Jpn\textbf{77}, 113712 (2008).
\bibitem{14} P. Alireza, Y. Ko, J. Gillett,
C. Petrone, J. Cole, G. Lonzarich and S. Sebastian1,
J.Phys.:Condens. Matter \textbf{21}, 012208 (2008).
\bibitem{Canfield}N. Ni, S. L. Bud'ko, A. Kreyssig, S. Nandi, G. E. Rustan, A. I. Goldman, S. Gupta, J. D. Corbett, A. Kracher, P. C.
Canfield, Phys. Rev. B \textbf{78}, 014507 (2008).
\bibitem{ChuCW2}K. Sasmal et al., Phys. Rev. Lett. \textbf{101}, 107007
(2008).
\bibitem{Mandrus}A. S. Sefat, R. Jin, M. A. McGuire, B. C. Sales, D. J. Singh, and D. Mandrus, Phys. Rev. Lett. \textbf{101}, 117004 (2008).
\bibitem{Fisher}J. H. Chu, J. G. Analytis, C. Kucharczyk and I. R. Fisher, Phys. Rev. B \textbf{79}, 014506 (2009).
\bibitem{FeP}Y. Kamihara, H. Hiramatsu, M. Hirano, R. Kawamura, H.
Yanagi, T. Kamiya, and H. Hosono, J. Am. Chem. Soc., {\bf128}, 10012
(2006).
\bibitem{FeP42622} H. Ogino, YutakaMatsumura, Y. Katsura,
K. Ushiyama, S. Horii, KohjiKishio and J. Shimoyama, Supercond. Sci.
Technol.,{\bf22}, 075008 (2009).
\bibitem{FeAs32522} X. Zhu, F. Han, G. Mu, P. Cheng, B. Shen, B. Zeng, and H. H. Wen, Phys. Rev. B \textbf{79}, 024516
(2009).
\bibitem{GFChen} G. F. Chen, T. L. Xia, P. Zheng, J. L. Luo, and N. L. Wang, arXiv:condmat/0903.5273 (2009).
\bibitem{XYZhu2} X. Zhu, F. Han, G. Mu, P. Cheng, B. Shen, B. Zeng, and H. H. Wen, arXiv:condmat/0904.0972 (2009).
\bibitem{Ogino} H. Ogino, Y. Katsura, S. Horii, K. Kishio, J. Shimoyama, arXiv: Condmat/0903.5124.
\bibitem{Johrendt}M. Tegel, F. Hummel, S. Lackner, I. Schellenberg, R. Poettgen, D. Johrendt, arXiv: Condmat/0904.0479.
\bibitem{ZhuXY}  X. Zhu, H. Yang, L. Fang, G. Mu, and
H. H. Wen, Supercond. Sci. Technol. \textbf{21}, 105001 (2008).
\bibitem{GSAS}   AC Larson, RB Von Dreele, General Structure Analysis System (GSAS),
Los Alamos National Laboratory Report LAUR 86-748, 2000.
\bibitem{WenEPL} H. H. Wen, G. Mu, L. Fang, H. Yang, and X. Zhu, Europhys. Lett. \textbf{82}, 17009 (2008).
\bibitem{MuGPRBPrSr} G. Mu, B. Zeng, X. Zhu, F. Han, P. Cheng, B. Shen, H. H. Wen, Phys. Rev. B \textbf{79}, 104501 (2009).
\bibitem{ZSWang} Z. S. Wang, H. Q. Luo, C. Ren, H. H. Wen, Phys. Rev. B \textbf{78}, 140501(R)
(2008).
\bibitem{WHH} N. R. Werthamer, E. Helfand, P. C. Hohenberg,  Phys. Rev. \textbf{147}, 295
(1966).
\bibitem{YangHuanPRL2008}H. Yang, Y. Liu, C. G. Zhuang, J. R. Shi, Y. G. Yao, S. Massidda, M. Monni, Y. Jia, X. X. Xi, Q. Li, Z. K. Liu, Q. R. Feng, H. H. Wen, Phys. Rev. Lett. {\bf101}, 067001 (2008).
\bibitem{FangL} L. Fang, H. Luo, P. Cheng, Z. Wang, Y. Jia, G. Mu, B. Shen, I. I. Mazin, L. Shan, C. Ren, H. H. Wen, arXiv: Condmat/0903.2418.
\end{thebibliography}
\end{document}